\documentclass[aps,pra,reprint,amsmath,amsfonts,amssymb,superscriptaddress]{revtex4-1}
\usepackage{graphicx,float,calc}
\usepackage{color,bm}
\usepackage{ulem}
\usepackage{braket}
\usepackage[colorlinks,urlcolor=blue,citecolor=blue,linkcolor=blue]{hyperref}

\begin{document}

\title{Interplay of nonreciprocity and nonlinearity on mean-field energy and dynamics of a Bose-Einstein condensate in a double-well potential}
\author{Yi-Piao Wu}
\affiliation{Guangdong Provincial Key Laboratory of Quantum Engineering and Quantum Materials, School of Physics and Telecommunication Engineering, South China Normal University, Guangzhou 510006, China}
\author{Guo-Qing Zhang}
\affiliation{Guangdong Provincial Key Laboratory of Quantum Engineering and Quantum Materials, School of Physics and Telecommunication Engineering, South China Normal University, Guangzhou 510006, China}
\affiliation{Guangdong-Hong Kong Joint Laboratory of Quantum Matter, Frontier Research Institute for Physics, South China Normal University, Guangzhou 510006, China}
\author{Cai-Xia Zhang}\thanks{zhangcx298@126.com}
\affiliation{Guangdong Provincial Key Laboratory of Quantum Engineering and Quantum Materials, School of Physics and Telecommunication Engineering, South China Normal University, Guangzhou 510006, China}
\affiliation{Guangdong-Hong Kong Joint Laboratory of Quantum Matter, Frontier Research Institute for Physics, South China Normal University, Guangzhou 510006, China}
\author{Jian Xu}\thanks{xujian\_328@163.com}
\affiliation{College of Electronics and Information Engineering, Guangdong Ocean University, Zhanjiang 524088, China}
\author{Dan-Wei Zhang}\thanks{danweizhang@m.scnu.edu.cn}
\affiliation{Guangdong Provincial Key Laboratory of Quantum Engineering and Quantum Materials, School of Physics and Telecommunication Engineering, South China Normal University, Guangzhou 510006, China}
\affiliation{Guangdong-Hong Kong Joint Laboratory of Quantum Matter, Frontier Research Institute for Physics, South China Normal University, Guangzhou 510006, China}

\begin{abstract}
We investigate the mean-field energy spectrum and dynamics in a Bose-Einstein condensate in a double-well potential with non-Hermiticity from the nonreciprocal hopping, and show that the interplay of nonreciprocity and nonlinearity leads to exotic properties. Under the two-mode and mean-field approximations, the nonreciprocal generalization of the nonlinear Schr\"{o}dinger equation and Bloch equations of motion for this system are obtained. We analyze the $\mathcal{PT}$ phase diagram and the dynamical stability of fixed points. The reentrance of $\mathcal{PT}$-symmetric phase and the reformation of stable fixed points with increasing the nonreciprocity parameter are found. Besides, we uncover a linear self-trapping effect induced by the nonreciprocity. In the nonlinear case, the self-trapping oscillation is enhanced by the nonreciprocity and then collapses in the $\mathcal{PT}$-broken phase, and can finally be recovered in the reentrant $\mathcal{PT}$-symmetric phase.	
\end{abstract}

\date{\today}

\maketitle

\section{Introduction}

Recent studies have demonstrated that non-Hermitian systems have intriguing physics and potential applications beyond Hermitian systems \cite{Bender_2007, ElGanainy2018,Miri2019,Ashida2020,Bergholtz2021}. Of particular interest is that non-Hermitian Hamiltonians subjecting to balanced gain and loss have the parity-time ($\mathcal{PT}$) symmetry that guarantees an purely real energy spectrum \cite{Bender1998}. The transitions between $\mathcal{PT}$-symmetric phases and $\mathcal{PT}$-broken phases happen with the spectral degeneracy, which is known as the exceptional point (EP) \cite{Heiss2004, Heiss2012, Pan2019}. Circling around an EP in the complex energy plane may introduce topological invariants unique to non-Hermitian Hamiltonians \cite{Gong2018, Kawabata2019}. Apart from the gain-and-loss effect, the nonreciprocity is another kind of non-Hermiticity. It has been theoretically revealed that nonreciprocal systems exhibit exotic topological and localization physics \cite{Hatano1996,Ashida2020,Bergholtz2021,Yao2018, Kunst2018, LJin2019, Longhi2019, HJiang2019, DWZhang2019,XWLuo2019,LZTang2020,HLiu2020,QBZeng2020, Li2020, Zhang2020,TLiu2020,ZXu2020}. In experiments, the nonreciprocity can be realized in classical electric circuits \cite{Helbig2020,Ezawa2019}, optical systems \cite{LXiao2020,Yu2009,Kang2011,Bi2011,Fan2011,Horsley2013,Peng2014, Wang2019, Zhao2020}, and optomechanical devices \cite{Shen2016, Ruesink2016, Bernier2017, Fang2017, Peterson2017, Xu_2019}. Remarkably, the tunable nonreciprocal hopping has recently been realized for an atomic condensate in a synthetic momentum lattice \cite{WGou2020}, which can also be engineered in optical lattices \cite{Gong2018,DWZhang2018}.

The nonlinearity is another important element that can induce a rich variety of novel phenomena in classical and quantum systems \cite{Kartashov2011}. For instance, the mean-field dynamics of a Bose-Einstein condensate in a double-well potential can be described by the nonlinear Schr\"{o}dinger and Bloch equations \cite{Morsch2006, Wu2000, Liu2002, Liu2003,Kellman2002, Hines2005, Siddiqi2005, Zibold2010,Lee2008,Lee2009,Lee2012,Modugno2017,Modugno2018,DWZhang2012,DWZhang2011,WYWang2021}. The nonlinear interaction therein can lead to exceptional crossing scenarios in the energy spectrum \cite{Wu2000, Liu2002, Liu2003} and the dynamical bifurcation \cite{Kellman2002, Hines2005, Siddiqi2005, Zibold2010,Lee2008,Lee2009}. The nonlinearity also enriches the Josephson dynamics of the atomic condensate in the double well, including the anharmonic Josephson oscillation and the macroscopical self-trapping effect with locked population imbalance between two wells \cite{Smerzi1997, Raghavan1999, Albiez2005, Abbarchi2013}. In recent years, there is growing attention devoted to investigate nonlinear properties of non-Hermitian systems \cite{Konotop2017}. Remarkably, the nonlinear mean-field dynamics in the non-Hermitian Bose-Hubbard dimer with dissipations \cite{Graefe2008, Graefe2008a, Witthaut2009,Graefe2013} and gain and loss \cite{Graefe2010, Graefe2012, Cartarius2012, Dast2012, Dast2013, Single2014, Fortanier2014, Dast2016, Haag2018} has been studied. The asymmetric loop spectra and symmetry breaking of the nonlinear Bloch spectrum in a $\mathcal{PT}$-symmetric periodic potential have recently been revealed \cite{YPZhnag2021}. Hitherto, the interplay of the nonreciprocity and nonlinearity and the associated effects in non-Hermitian systems are largely unexplored.

Inspired by the experimental realizations of nonreciprocal hopping and double-well potentials for Bose-Einstein condensates (BECs), in this work, we explore the mean-field energy spectrum and dynamics of a BEC with nonlinear interactions and nonreciprocal hopping in a double-well potential. Based on the two-mode and mean-field approximations, we obtain the nonreciprocal generalization of the nonlinear Schr\"{o}dinger equation and Bloch equations of motion for the system. The $\mathcal{PT}$ phase diagram and the corresponding dynamical stability of fixed points are analyzed. Furthermore, we demonstrate that the interplay of nonreciprocity and nonlinearity leads to several novel and unique properties in the energy spectrum and dynamics: i) We find the reentrance of $\mathcal{PT}$-symmetric phase and the reformation of stable fixed points with increasing the nonreciprocity parameter for large nonlinear interactions. ii) We uncover a new self-trapping effect solely induced by the nonreciprocity in the linear case. iii) In the nonlinear case, the self-trapping oscillation of the population imbalance is enhanced by the nonreciprocity and then collapses in the $\mathcal{PT}$-broken phase, and can finally be recovered in the reentrant $\mathcal{PT}$-symmetric phase. The predicted features may be tested in future experiments with ultracold bosonic atoms since all the required ingredients, which include tunable nonreciprocal hopping, double-well potential and interactions for atomic condensates, have already been experimentally achieved.

The rest of this article is organized as follows. Section \ref{sec2} introduces the model and equations of motions. In Sec. \ref{sec3}, we present the results of the $\mathcal{PT}$ phase diagram for mean-field eigenenergies, the dynamical stability of fixed points, and the self-trapping and reentrant oscillation effects. A brief discussion on the renormalization condition and a short summary are finally given in Sec. \ref{sec4}.

\section{\label{sec2}MODEL}

We consider a BEC in a symmetric double-well trapping potential with $N$ weakly interacting atoms, and assume that the atoms are coherently transferred from one well to the other with laser-induced tunable nonreciprocal hopping \cite{WGou2020}. Within the two-mode approximation \cite{Smerzi1997, Raghavan1999, Albiez2005, Abbarchi2013,Lee2008,Lee2009,Lee2012,Modugno2017,Modugno2018}, the system can be described by the following two-site Bose-Hubbard Hamiltonian with the nonreciprocal hopping \cite{Yao2018, Kunst2018, LJin2019, Longhi2019, HJiang2019, DWZhang2019,XWLuo2019,LZTang2020,HLiu2020,QBZeng2020, Li2020, Zhang2020,TLiu2020,ZXu2020}:
\begin{equation}\label{Ham}
	\begin{split}
		\hat{\mathcal{H}}_{BH}=-&J\hat{a}^\dag_1 \hat{a}_2-J(1-\gamma)\hat{a}^\dag_2 \hat{a}_1 +\frac U2 (\hat{a}^\dag_1 \hat{a}_1-\hat{a}^\dag_2 \hat{a}_2)^2,
	\end{split}
\end{equation}
where $\hat{a}_l$ and $\hat{a}^\dag_l$ ($l=1,2$) are bosonic annihilation and creation operators
on the $l$-th well (mode), $J$ is the hopping constant, $\gamma\geqslant0$ denotes the nonreciprocity parameter, $U$ is the on-site repulsive interaction strength. For convenience, we set reduced Planck constant $\hbar =1$ and $J=1$ as the energy unit hereafter. The Hamiltonian (\ref{Ham}) satisfies the $\mathcal{PT}$-symmetry, i.e., $[\mathcal{\hat{P}\hat{T}}, \hat{\mathcal{H}}_{BH}]=0$, where $\mathcal{\hat{P}}$ is the parity operator acting as the inversion operator $\hat{a}_1 \leftrightarrow \hat{a}_{2}$ and $\mathcal{\hat{T}}$ is time reversal operator acting as the complex conjugation $\hat{a}^\dag_l \leftrightarrow \hat{a}_l$. In the following, we focus on the non-Hermitian effects on the mean-field energy and dynamics.

In the limit of $N\rightarrow\infty$, one can obtain the semiclassical counterpart of the second quantized Hamiltonian by the mean-field approximation. In this case, the annihilation and creation operators can be replaced by their expectation values with
complex numbers: $\hat{a}_l\approx\langle\hat{a}_l\rangle\equiv\psi_l$ and $\hat{a}^\dag_l\approx\langle\hat{a}^\dag_l\rangle\equiv\psi^\ast_l$. Below we will begin on the mean-field model Hamiltonian for the macroscopic limit $N\rightarrow\infty$ and nonlinear interaction strength $g=NU$. We use the rescaled mean-field wave function $\psi_l\rightarrow \sqrt N \psi_l$ and focus on the normalized condition $|\psi_1|^2+|\psi_2|^2=1$. Then we obtain the energy functional of the Hamiltonian (\ref{Ham}) as
\begin{equation}\label{Ham1}
	\begin{split}
		\mathcal{E}=-J\psi^\ast_1 \psi_2-J(1-\gamma)\psi^\ast_2 \psi_1 +\frac{g}{2}(\psi^\ast_1\psi_1-\psi^\ast_2\psi_2)^2,
	\end{split}
\end{equation}
The time evolution of the mean-field wave function can be obtained by $i\frac{\partial\psi_l}{\partial t}=\frac{\partial \mathcal{E}}{\partial \psi^\ast_l}$ \cite{WYWang2021}. This leads to the following nonreciprocal version of the nonlinear Schr\"{o}dinger equation, i.e., the Gross-Pitaevskii (GP) equation:
\begin{equation}\label{GPE}
	\begin{aligned}
		i \frac{d}{dt}
		\begin{pmatrix}
			\psi_1 \\
			\psi_2 \\
		\end{pmatrix}
	=\hat{\mathcal{H}}_{\text{GP}}
		\begin{pmatrix}
			\psi_1 \\
			\psi_2 \\
		\end{pmatrix}
       =
		\begin{pmatrix}
			g\kappa& -J \\
			-(1-\gamma)J & -g\kappa
		\end{pmatrix}
		\begin{pmatrix}
			\psi_1 \\
			\psi_2 \\
		\end{pmatrix}
	\end{aligned}
\end{equation}
where $\kappa=|\psi_1|^2-|\psi_2|^2$ is the population imbalance. By defining the normalized Bloch vectors $s_j$ with $j=x,y,z$ from the wave functions $\psi_l$: $s_x=\frac{1}{2}(\psi^*_1\psi_2+\psi_1\psi^*_2)$, $s_y=\frac{1}{2i}(\psi^*_1\psi_2-\psi_1\psi^*_2)$, and $s_z=\frac{1}{2}(\psi^*_1\psi_1-\psi_2\psi^*_2)=\kappa/2$, we can obtain the mean-field Bloch equations of motion:
\begin{equation}\label{EoM2}
	\begin{aligned}
		\dot{s}_x&=-4g s_z s_y +2\gamma Js_y s_x,\\
		\dot{s}_y&=4g s_z s_x +(2-\gamma)Js_z +2\gamma Js^2_y -\frac12 \gamma J,\\
		\dot{s}_z&=-(2-\gamma)Js_y +2\gamma J s_z s_y.
	\end{aligned}
\end{equation}
These nonlinear and nonreciprocal Bloch equations take real values and conserve as $s_x^2+s_y^2+s_z^2=1/4$. Accordingly, the motion of the Bloch vectors under the renormalization is described by the trajectories on the surface of the Bloch sphere. At the end of this paper, we will consider the Bloch dynamics without renormalization and show that our results in the following section preserves.

\begin{figure}
	\centering
	\includegraphics[width=0.45\textwidth]{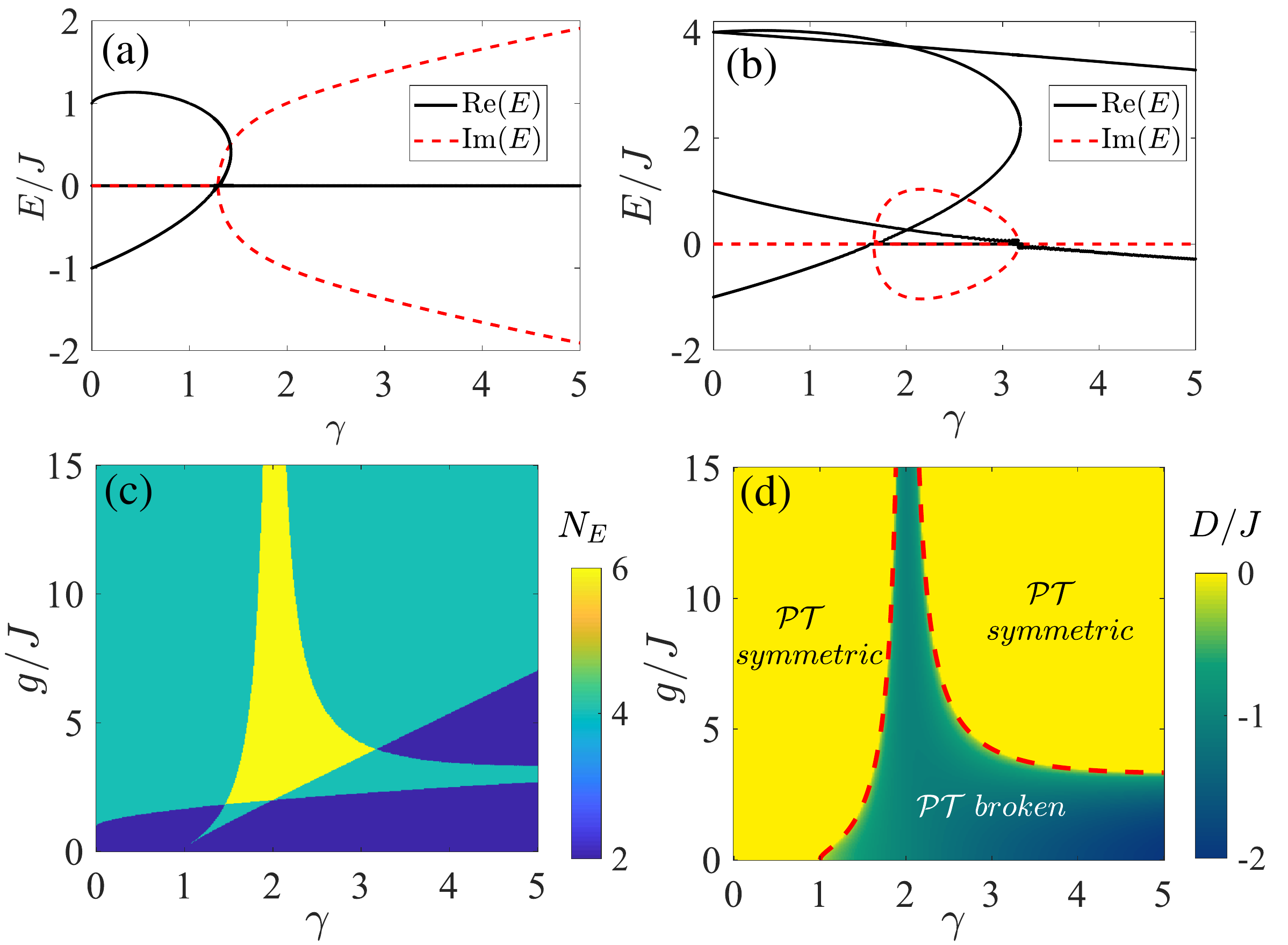}
	\caption{Mean-field energy spectra as a function of the nonreciprocity $\gamma$ for the nonlinear strength $g = 1$ (a) and $g = 4$ (b). (c) Number of eigenenergies $N_E$ as functions of $\gamma$ and $g$. (d) $\mathcal{PT}$ phase diagram on the $\gamma$-$g$ plane determined by the maximum imaginary energy $D$. The $\mathcal{PT}$-symmetric and $\mathcal{PT}$-broken phases are labeled. The red dashed lines denote the analytical results for the critical boundaries given by Eq. (\ref{boundary}), which agree well with the numerical results.}
\label{fig1}
\end{figure}

\section{\label{sec3} Results}

In this section, we study the interplay of nonreciprocity and nonlinearity in the mean-field energy spectrum and dynamics. We present the analysis and discussion on the numerical results.


\begin{figure}
	\centering
	\includegraphics[width=0.4\textwidth]{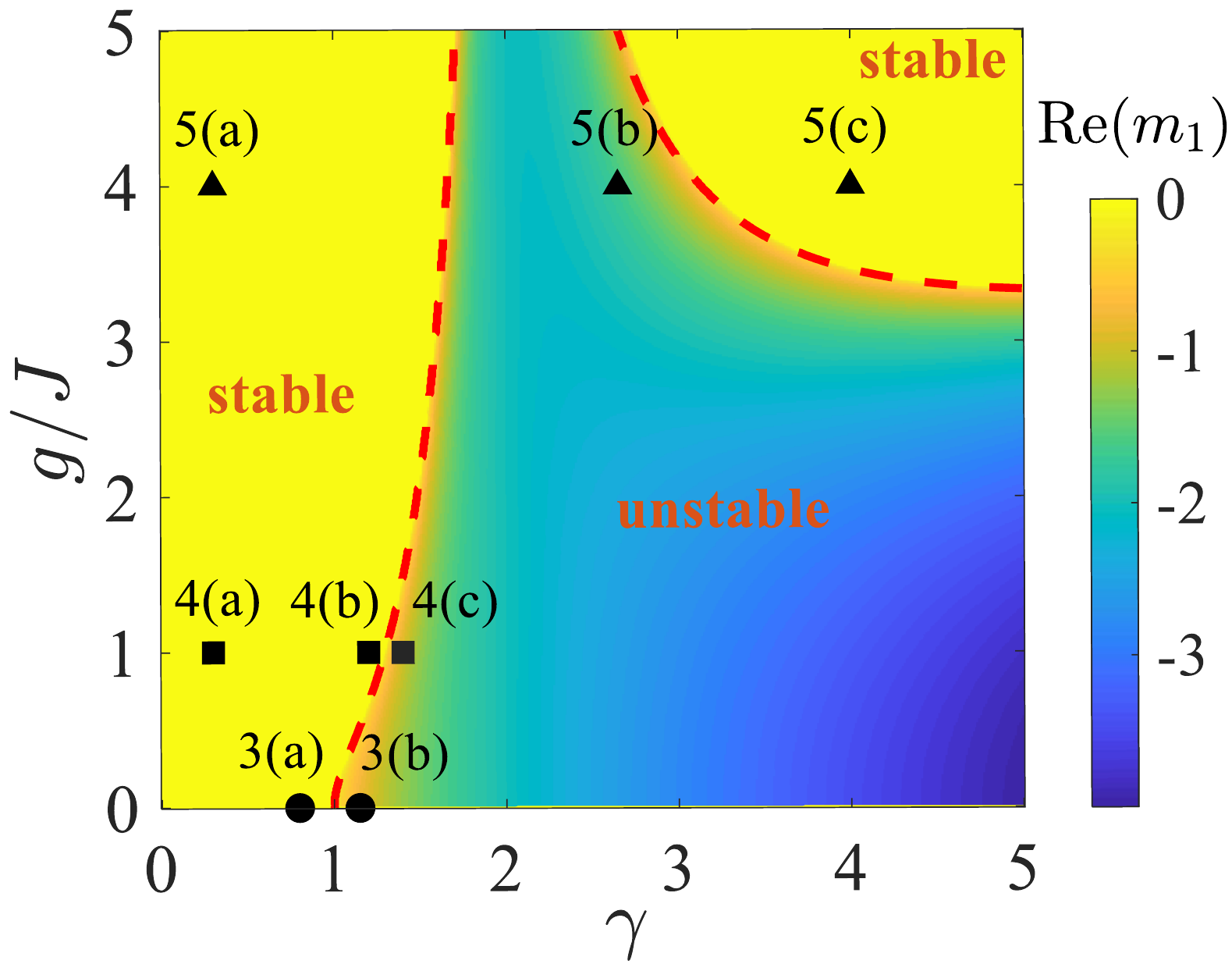}
	\caption{Dynamical stability diagram for $J_{f_{-}}$ fixed points of the Bloch equations (\ref{EoM2}). The $\gamma$-$g$ parameter regions for the center and unstable fixed points are labeled by $\text{Re}(m_1) = 0$ and $\text{Re}(m_1) > 0$, respectively. The red dashed lines correspond to the boundaries of $\mathcal{PT}$ symmetric and broken phases shown in Fig. \ref{fig1}(d). The labels denotes the parameters for the resulting Bloch dynamics shown in Figs. (\ref{fig3}-\ref{fig5}).
	}\label{fig2}
\end{figure}


\subsection{$\mathcal{PT}$ phase diagram for mean-field eigenenergies}

Before considering the Bloch dynamics, we first study the mean-field energy spectrum by
numerically solving the time-independent GP equation
\begin{equation}\label{GPE2}
	\begin{aligned}
     \hat{\mathcal{H}}_{\text{GP}}
		\begin{pmatrix}
			\psi_1 \\
			\psi_2 \\
		\end{pmatrix}
		=E
		\begin{pmatrix}
			\psi_1 \\
			\psi_2 \\
		\end{pmatrix}
	\end{aligned}.
\end{equation}
Here eigenstates $\psi_l$ and eigenenergy $E$ are self-consistent solutions for given $\gamma$ and $g$. In the Hermitian limit with $\gamma=0$, there are 2 or 4 real eigenenergies for the two-mode GP Hamiltonian for $g\leqslant1$ or $g>1$ \cite{Wu2000, Liu2002, Liu2003}, respectively.


\begin{figure}
	\centering
	\includegraphics[width=0.45\textwidth]{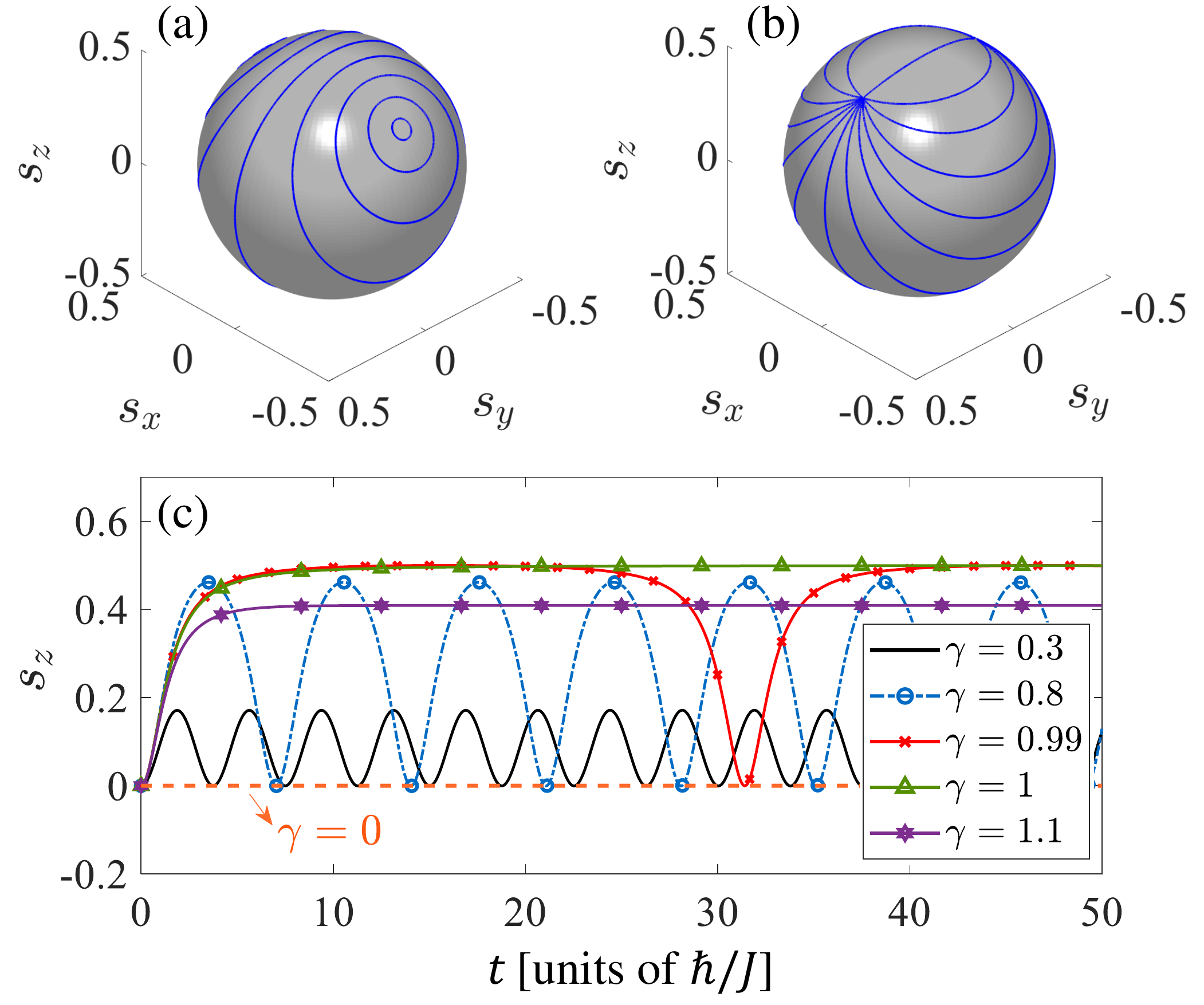}
	\caption{Mean field dynamics in the linear case $g=0$. The dynamics of the mean-field Bloch vector on the Bloch sphere for the nonreciprocity parameter $\gamma=0.8$ (a) and $\gamma=1.1$ (b). The blue curves denotes the resulting dynamical trajectories after a sufficiently long evolution time. (c) Time evolution of the population
imbalance $s_z(t)$ for various $\gamma$ under the initial condition $s_z(0)=(1/2,0,0)$.
	}\label{fig3}
\end{figure}


For our nonreciprocal GP Hamiltonian, we numerically find that the eigenenergy number $N_E$ can be 2, 4, or 6, which is dependent on $g$ and $\gamma$ [see Fig. \ref{fig1}(c)]. Two typical energy spectrums with the real part $\text{Re}(E)$ and the imaginary part $\text{Im}(E)$ with respect to the nonreciprocity $\gamma$ for $g=1$ and $g=4$ are plotted in Figs. \ref{fig1}(a) and \ref{fig1}(b), respectively. For $g=1$, the eigenenergies remain real with increasing $\gamma$ up to the EP, i.e., $\gamma<\gamma_{\text{EP}}\approx1.27$, which is denoted as a  $\mathcal{PT}$-symmetric phase \cite{Bender1998, Bender_2007, ElGanainy2018}. The system undergoes a spontaneous $\mathcal{PT}$-symmetry-breaking transition at the EP $\gamma=\gamma_{\text{EP}}$, where two of four real eigenenergies become degenerate and their corresponding eigenvectors coalesce simultaneously, and then enters the $\mathcal{PT}$-broken phase with $\text{Im}(E)\neq0$ when $\gamma>\gamma_{\text{EP}}$. For larger nonlinearity $g=4$ [Fig. \ref{fig1}(b)], the system becomes $\mathcal{PT}$ broken after the first EP $\gamma_{\text{EP1}}\approx1.62$. Moreover, the system will reenter the $\mathcal{PT}$-symmetric phase with further increasing the nonreciprocity $\gamma$ after the second EP $\gamma_{\text{EP2}}\approx3.17$. We further display the eigenenergy number $N_E$ and the $\mathcal{PT}$ phase diagram on the $\gamma$-$g$ plane in Figs. \ref{fig1}(c) and \ref{fig1}(d), respectively. Here the maximum value of the imaginary parts of all eigenvalues $D=\text{Max}[|\text{Im}(E_m)|]$ with $m=1,2,...,N_E$ is used to seperate the $\mathcal{PT}$-symmetric phase with $D=0$ and the $\mathcal{PT}$-broken phase with $D>0$. As we will see in the following section, the renormalized population dynamics generally exhibit the oscillation (nonoscillational decay) behaviour in the $\mathcal{PT}$-symmetric ($\mathcal{PT}$-broken) phase. From Fig. \ref{fig1}(d), one can find that the energy spectrum of our model is real when $\gamma<1$ and has the $\mathcal{PT}$ breaking transition at $\gamma=1$ in the linear case with $g=0$. Turning up the nonlinear interaction will enlarge the $\mathcal{PT}$-phase region. Most strikingly, we find the reentrance of $\mathcal{PT}$-symmetric phase with increasing the nonreciprocity parameter $\gamma$ when $g>g_c\approx3.34$.

\begin{figure*}
	\centering
	\includegraphics[width=0.75\textwidth]{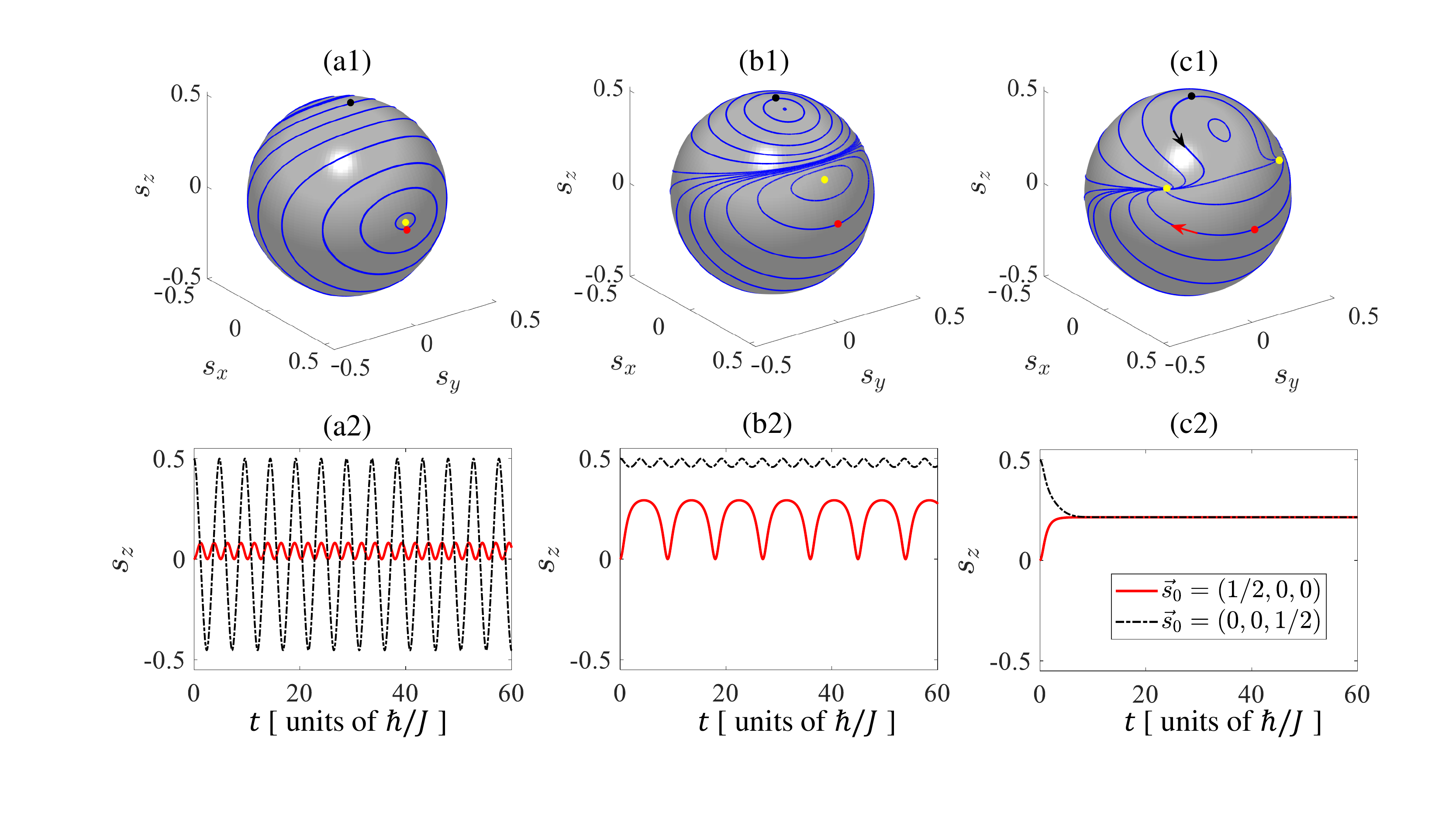}
	\caption{Mean-field dynamics for $\gamma=0.3$ (a), $\gamma=1.2$ (b), and $\gamma=1.4$ (c) with fixed $g=1$. The upper figures show the resulting dynamical trajectories on the Bloch sphere after a sufficiently long evolution time. The lower figures show the corresponding time evolution of the population imbalance $s_z(t)$ under the initial condition $\vec{s}_0(0)=(1/2,0,0)$ (red solid lines) and $\vec{s}_0(0)=(0,0,1/2)$ (black dashed lines). Certain fixed points and the two initial states are marked by yellow, red and black dots on the Bloch sphere, respectively. The arrows in (c1) denote the direction of the trajectory flow.
	}\label{fig4}
\end{figure*}

To confirm our numerical results of the $\mathcal{PT}$ phase diagram, we give the form solutions of the GP equation (\ref{GPE2}) with the corresponding eigenenergies and eigenstates:
\begin{equation}\label{solutions}
	\begin{aligned}
E=&\pm \sqrt{F(\gamma,g)}= \pm\sqrt{g^2\kappa^2+(1-\gamma)J^2},\\
\begin{pmatrix}
	\varphi_1 \\
	\varphi_2 \\
\end{pmatrix}
=&\frac{1}{\sqrt{1+\alpha^2}}
\begin{pmatrix}
	1 \\
	\alpha \\
\end{pmatrix},
	\end{aligned}
\end{equation}
where $\alpha=-(E-g\kappa)/J$. For the EPs at $E=0$ and the $\mathcal{PT}$-broken phase with imaginary eigenenergies [see Figs. \ref{fig1}(a) and \ref{fig1}(b) for examples], one has the solutions satisfying $E^*=-E$, which leads to the simplified imbalance parameter $\kappa=(1-\alpha^2)/(1+\alpha^2)=(2-\gamma)/\gamma$. In these cases, the self-consistent solutions require  $F(\gamma,g)=0$ for EPs and $F(\gamma,g)<0$ for the $\mathcal{PT}$-broken phase region in the parameter space. Thus, we obtain the critical condition equation
\begin{equation}\label{boundary}
(2-\gamma)^2 g^2/J^2+\gamma^2-\gamma^3=0,
\end{equation}
which determines the $\mathcal{PT}$-symmetry breaking. The resulting phase boundaries on the $\gamma$-$g$ plane are plotted as the red dashed lines in Fig. \ref{fig1}(d), which agree well with our numerical results.

The $\mathcal{PT}$ phase diagram can also be intuitively understood by rewriting the nonreciprocal GP Hamiltonian as
\begin{equation}\label{GPE3}
\hat{\mathcal{H}}_{\text{GP}}=g\kappa\hat{\sigma}_z-(1-\frac{\gamma}{2})J\hat{\sigma}_x-i\frac{\gamma}{2}J\hat{\sigma}_y,  \end{equation}
where $\hat{\sigma}_{x,y,z}$ are the Pauli matrices acting on two modes. Here the first term is the Hermitian nonlinear interaction, the second term is the coherent hopping, and the third term represents the non-Hermitian and incoherent hopping. The interplay among the three terms gives rise to real-complex eigenenergy transitions shown in Fig. \ref{fig1}(d). For $g=0$, the energy spectrum is real (complex) when $\gamma<1$ ($\gamma>1$) as the coherent (incoherent) hopping dominates and has the $\mathcal{PT}$-symmetric breaking transition at $\gamma=1$. Increasing $g$ can then enlarge the $\mathcal{PT}$ phase region when $\gamma<2$. At the critical case of $\gamma=2$, the coherent hopping strength ($1-\gamma/2=0$) and population imbalance ($\kappa=|\psi_1|^2-|\psi_2|^2=0$ verified in our numerical simulations) are vanishing, such that the system always stays in $\mathcal{PT}$-broken phase for any finite $g$. When $\gamma\gg2$, the coherent and incoherent hopping terms are close, and then a sufficiently large interaction ($g>g_c\approx3.34$) will lead to the real energy spectrum. Thus, the reentrant $\mathcal{PT}$-symmetric phase exhibits in the parameter region of large $g$ and $\gamma$.


\begin{figure*}
\centering
\includegraphics[width=0.75\textwidth]{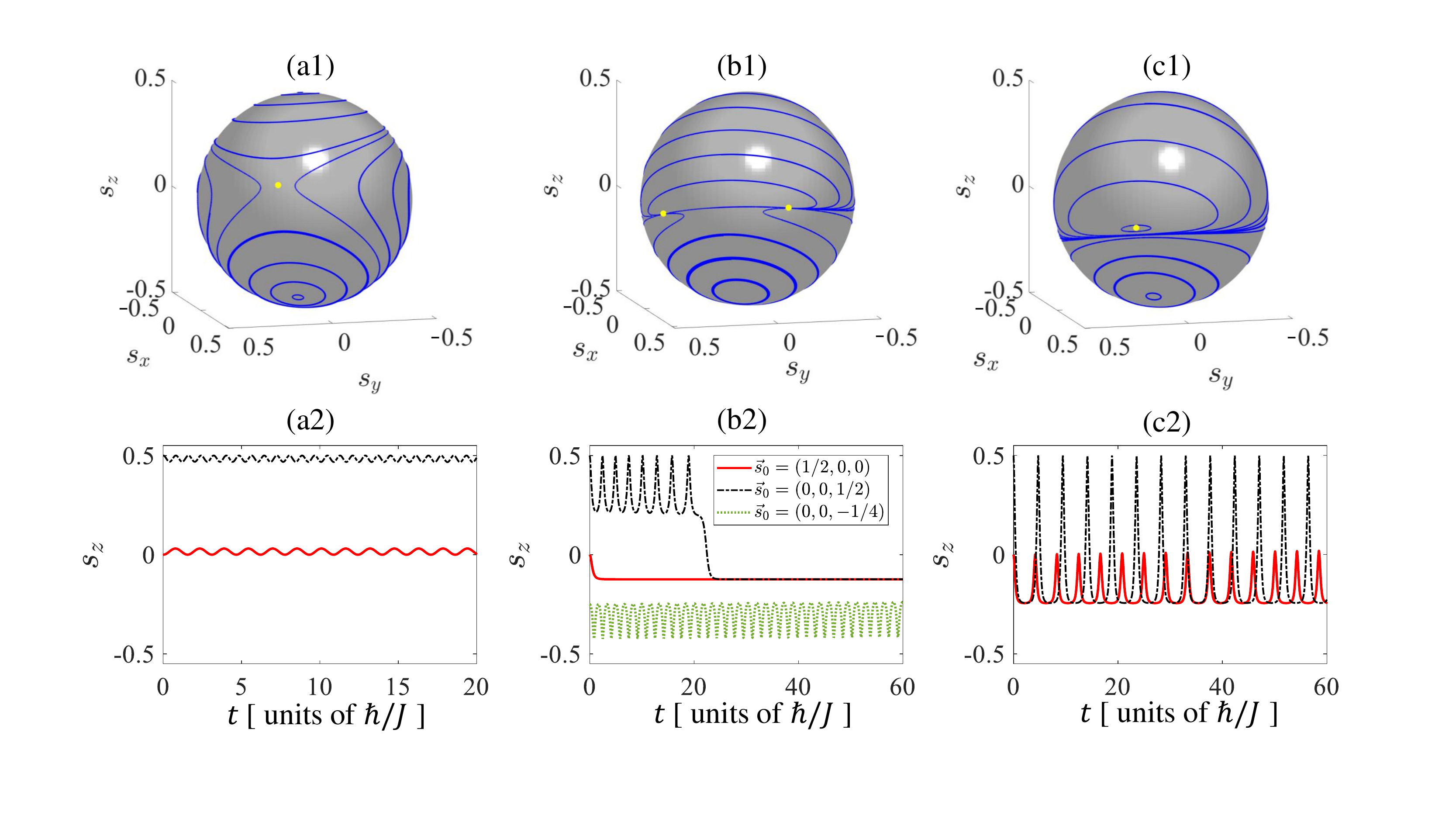}
\caption{Mean-field dynamics for $\gamma=0.3$ (a), $\gamma=2.66$ (b), and $\gamma=4$ (c) with fixed $g=4$. The upper figures show the resulting dynamical trajectories on the Bloch sphere. The lower figures show the corresponding time evolution of the population imbalance $s_z(t)$ under the initial condition $\vec{s}_0(0)=(1/2,0,0)$ (red solid lines), $\vec{s}_0(0)=(0,0,1/2)$ (black dashed lines), and $\vec{s}_0(0)=(0,0,-1/4)$ (green dotted line). Certain fixed points are marked by yellow dots on the Bloch sphere.
}\label{fig5}
\end{figure*}


\subsection{Dynamical stability of fixed points}

We now investigate the mean-field dynamics arising from the interplay of the nonreciprocity and nonlinearity. The
dynamics described by Eq. (\ref{EoM2}) can be understood as quantum trajectories with fixed points on the Bloch sphere. The fixed points are singular points of the Bloch vector field, which correspond to stationary solutions of the nonlinear Bloch equations. However, stationary solutions can be dynamically unstable in certain parameter space under arbitrary small perturbations to fixed points on the Bloch sphere \cite{BWu2001, BWu2003}. For our model, we first analyze the dynamical instabilities of fixed points on the Bloch sphere under the interplay of nonreciprocity and nonlinearity. By solving the fixed point equation defined by Eq. (\ref{EoM2}) with $\dot{\vec{s}}\equiv(\dot{s}_x,\dot{s}_y,\dot{s}_z)^T=0$, we can obtain a pair of fixed points located at $\vec{f}_{\pm}=(f_x,\pm f_y,f_z)^T$ with
\begin{equation}\label{FP}
	\begin{aligned}
		f_x=&(2-\gamma)g/(\gamma^2J) \\
		f_y=&\sqrt{1-(2-\gamma)^2/\gamma^2-4g^2(2-\gamma)^2/(\gamma^4J^2)}/2 \\
		f_z=&(2-\gamma)/(2\gamma)
	\end{aligned}
\end{equation}
for given system parameters.

To study the dynamical stability of the fixed points, we can perform small perturbation to them with the linear treatment as \cite{Seyranian2003}
\begin{equation}\label{ }
	\delta \dot{\vec{f}}_{\pm}=J_{\vec{f}_{\pm}} \delta\vec{f}_{\pm},
\end{equation} 		
where the Jacobi matrix of fixed points is given by
\begin{equation}\label{Jacobi}
	J_{\vec{f}_{\pm}}=\left(
	\begin{matrix}
		\pm 2\gamma f_{y}  & -4gf_{z}                    & \mp 4gf_{y}            \\
		4gf_{z}         & \pm 4\gamma f_{y}              &  4gf_{x}+(2-\gamma) \\
		0                 & -(2-\gamma)+2\gamma f_{z}    & \pm 2\gamma f_{y}      \\
	\end{matrix}
	\right).
\end{equation}
The real part of the eigenvalues of the Jacobi matrices yield information about the type of the points \cite{Seyranian2003}. For the center, one has $\text{Re}(m_1)=\text{Re}(m_2)=\text{Re}(m_3)=0$, while $\text{Re}(m_l)> 0$ for the unstable fixed point and $\text{Re}(m_l)<0$ for the stable fixed point. We take the numerical result of $\text{Re}(m_1)$ of $J_{\vec{f}_{-}}$ to determine the unstable point $\vec{f}_{-}$ and to obtain the dynamical stability diagram on the $\gamma$-$g$ plane [also numerically confirmed by $\text{Re}(m_{2,3})$], as shown in Fig. \ref{fig2}.

The presence of unstable fixed points indicates the destruction of the periodic motion of Bloch vectors for certain initial conditions. Thus, the critical boundaries indicate the occurrence of fixed point bifurcations and the sharp charge of the resulting Bloch dynamics, as we will discuss in the following section. One can also find that the boundaries between the stable and unstable regions are consistent with those between $\mathcal{PT}$ symmetric and broken phases shown in Fig. \ref{fig1}(d). This indicates the correspondence of the complex eigenenergies and the unstable fixed points.

\subsection{Self-trapping and reentrant oscillations}

In the Hermitian and linear limit, i.e., $\gamma=g=0$, the mean-field dynamics in our system recover to the well-known Josephson oscillation of the population of each well \cite{Smerzi1997, Raghavan1999, Albiez2005, Abbarchi2013}. It is also well-known that the nonlinear interaction can lead to the anharmonic generalization of Josephson oscillations and the so-called self-trapping effect \cite{Smerzi1997, Raghavan1999, Albiez2005, Abbarchi2013}, that is, a self-locked population imbalance with a nonzero time averaged $\langle s_z(t)\rangle\neq0$. Here we reveal a different novel self-trapping effect solely induced by the nonreciprocal hopping in the linear case of $g=0$.

We show the numerical results of dynamical trajectories on the Bloch sphere after a sufficiently long evolution time for $\gamma=0.8$ and $\gamma=1.1$ with fixed $g=0$ in Figs. \ref{fig3}(a) and \ref{fig3}(b), respectively. When $0<\gamma<1$, we find a pair of stable fixed points as the oscillation centers localized at $s_z=f_z\neq0$ [Fig. \ref{fig3}(a)], which indicates the population in each well oscillates around $\langle s_z(t)\rangle\neq0$. To be more clearly, we plot the time evolution of the population imbalance $s_z(t)$ for various values of $\gamma$ under the initial condition $s_z(0)=(1/2,0,0)$ in Fig. \ref{fig3}(c). One can see that the self-trapping oscillation is turned on from $\gamma=0$ and exhibits larger amplitudes and longer periods with increasing $\gamma$ when $0<\gamma<1$. When $\gamma\geqslant1$, the two oscillation centers in the Bloch sphere become unstable nodes [Fig. \ref{fig3}(b)], which signals that the oscillatory behavior will break down. As expected, we find the nonoscillatory self-trapping of the population imbalance $s_z(t)$ for $\gamma\geqslant1$ in Fig. \ref{fig3}(c). Notably, the nonreciprocity-induced self-trapping effect without nonlinear interactions has not been found in the non-Hermitian dissipation \cite{Graefe2008, Graefe2008a, Witthaut2009} and gain-and-loss systems  \cite{Graefe2008, Graefe2008a, Witthaut2009, Graefe2010, Graefe2012, Cartarius2012, Dast2012, Dast2013, Single2014, Fortanier2014, Dast2016, Haag2018}.

We further study the interplay of the nonreciprocity and nonlinearity on the mean-field dynamics. The numerical results for $\gamma=0.3,1.2,1.4$ and fixed $g=1$ are shown in Fig. \ref{fig4}. In the $\mathcal{PT}$-symmetric phase [Figs. \ref{fig4}(a) and  \ref{fig4}(b)], the Bloch vectors exhibit period motion on the Bloch sphere and the population imbalance $s_z(t)$ exhibit the self-trapping oscillation. With increasing the nonreciprocity $\gamma$, the self-trapping oscillation is first promoted and then collapse after entering the $\mathcal{PT}$-broken phase [Fig. \ref{fig4}(c2)]. The collapse (nonoscillation) dynamics of the population imbalance correspond to the charging of oscillation centers into sink and source nodes of the dynamical trajectories on the Bloch sphere [Fig. \ref{fig4}(c1)].


\begin{figure}
	\centering
	\includegraphics[width=0.4\textwidth]{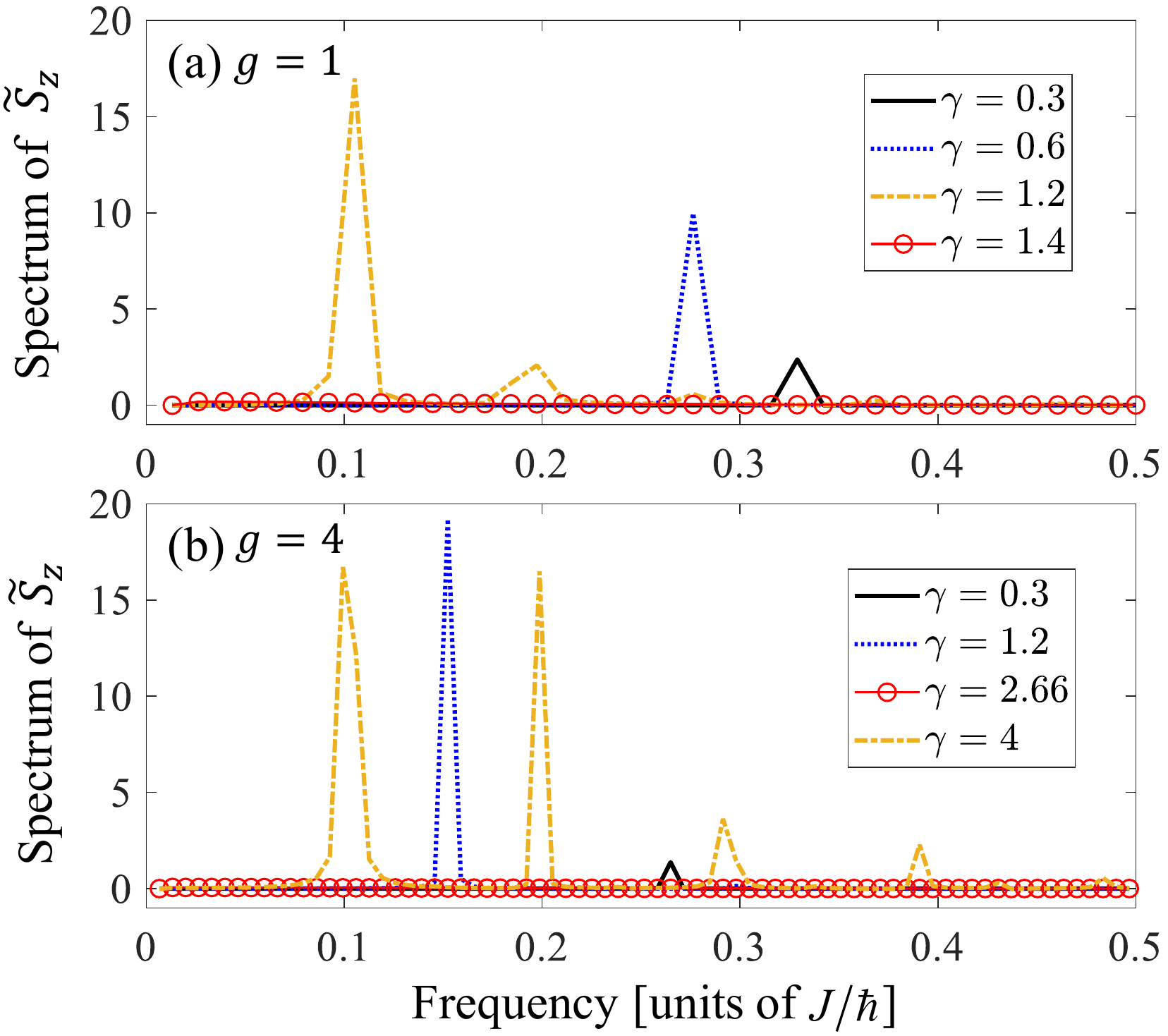}
	\caption{Frequency spectra of the relative population imbalance $\tilde{s}_z(t)$ for different $\gamma$ and fixed $g=1$ (a) and $g=4$ (b) under the initial condition $\vec{s}_0(0)=(1/2,0,0)$.
	}\label{fig6}
\end{figure}



\begin{figure*}
	\centering
	\includegraphics[width=0.75\textwidth]{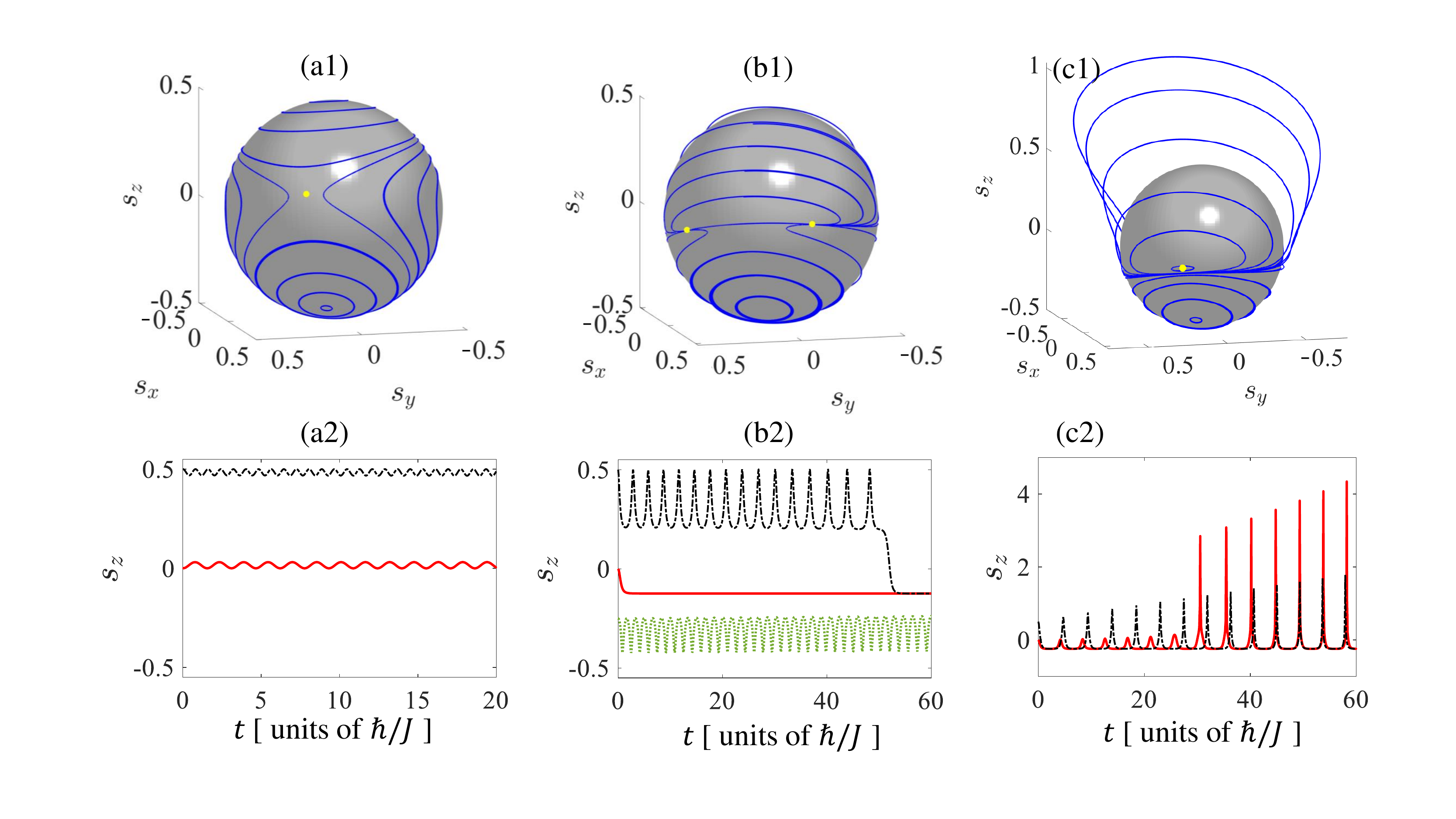}
	\caption{Non-renormalized mean-field dynamics for $\gamma=0.3$ (a), $\gamma=2.66$ (b), and $\gamma=4$ (c) with fixed $g=4$. All representations of dots and lines are the same to those in Fig. \ref{fig5}.}\label{fig7}
\end{figure*}


For large nonlinear interaction with $g>g_c\approx3.34$, we find that the self-trapping oscillation can be recovered from the collapse dynamics by further increasing the nonreciprocity due to the reentrance of $\mathcal{PT}$-symmetric phase. We show the numerical results for $\gamma=0.3,2.66,4$ and fixed $g=4$ in Fig. \ref{fig5}. For $\gamma = 0.3$ [Fig. \ref{fig5}(a)], there are mixed Bloch dynamics surrounding four fixed points, which contain a saddle point (marked by the yellow dot) and three centers (two of them are at the poles and the third one is localized at the equator opposite to the saddle point). The corresponding time evolutions of $s_z(t)$ for initial states $\vec{s}(0)=(0,0,1/2)$ and $\vec{s}(0)=(1/2,0,0)$ show two slight self-trapping oscillations. For the $\mathcal{PT}$-broken phase with $\gamma= 2.66$ [Fig. \ref{fig5}(b)], the two oscillations collapse after some evolution time. However, for proper initial states, such as $\vec{s}(0)=(-1/4,0,0)$, the oscillation can preserve all the time. This behaviour can be understood from the bifurcation of fixed points on the Bloch sphere \cite{Kellman2002, Hines2005, Siddiqi2005, Zibold2010}. As shown in Fig. \ref{fig5}(b1), the saddle point on Bloch sphere is bifurcated into a pair of fixed points, where the dynamical trajectories are terminated for certain initial states, such as those in the upper hemisphere. In these cases, the oscillation may last for a while but collapses after a long time. While for certain initial states in the lower hemisphere, the resulting dynamical trajectories surround the center at the south pole, which yields the period motion. By further increasing the nonreciprocity parameter to the reentrant $\mathcal{PT}$-symmetric phase, such as $\gamma=4$ [Fig. \ref{fig5}(c)], we can find that the paired fixed points collide and become a oscillation center in the Bloch sphere. Accordingly, the self-trapping oscillations of the population imbalance $s_z(t)$ recover with larger amplitudes. Thus, the collapse and revival of self-trapping oscillations is due to the interplay of nonreciprocity and nonlinearity.

We further calculate the frequency spectra of the relative population imbalance
\begin{equation}
	\begin{aligned}
\tilde{s}_z(t)\equiv s_z(t)-s_{zc}
	\end{aligned}
\end{equation}
via Fourier transform, where $s_{zc}$ denotes the center value and is determined by $s_{zc}=\{\text{Max}[s_z(t)]-\text{Min}[s_z(t)]\}/2$. As shown in Fig. \ref{fig6}, with the increase of nonreciprocity parameter $\gamma$, the oscillation frequency decreases and the amplitude increases initially. For $g=1$ in Fig. \ref{fig6}(a), more than one peak emerge and then disappear in the frequency spectra by further increasing $\gamma$. For $g=4$ in Fig. \ref{fig6}(b),
in contrast, the peak disappears and then more than one peak emerge in the frequency spectra. The multi-peaks in the spectra correspond to the nonsinusoidal self-trapping oscillations in Figs. \ref{fig4}(b2) and \ref{fig5}(c2). The disappearence and reappearence of peaks in the spectra correspond to collapse and revival of self-trapping oscillations, respectively.

\section{\label{sec4} DISCUSSION AND SUMMARY}

Before concluding, we emphasize that the novel self-trapping effects revealed previously are qualitatively independent on the renormalization condition. We numerically solve the generalized Bloch equations (\ref{EoM2}) without the renormalization condition $s_x^2+s_y^2+s_z^2=1/4$. For comparisons, we take the same parameters and initial conditions as those in Fig. \ref{fig5}, and show the results of non-renormalized mean-field dynamics in Fig. \ref{fig7}. One can find that in the $\mathcal{PT}$-broken phase for small $\gamma$ [Fig. \ref{fig7}(a)], the non-renormalized mean-field dynamics remains the same with that under the renormalization condition [Fig. \ref{fig5}(a)]. For intermediate and large $\gamma$ [Figs. \ref{fig7}(b) and \ref{fig7}(c)], although certain non-renormalized dynamical trajectories no longer stay on the Bloch sphere, the collapse and revival of self-trapping oscillations still exhibit with only quantitative differences. We also confirm similar comparison results between the non-renormalized and renormalized dynamics for other cases in Figs. \ref{fig3} and \ref{fig4}.

In summary, we have explored the interplay of nonreciprocal hopping and nonlinear interactions on the mean-field energy spectrum and dynamics of a Bose-Hubbard dimer model. The nonreciprocal generalizations of the nonlinear Bloch and Schr\"{o}dinger equations for the system have been obtained. We have found the reentrance of $\mathcal{PT}$-symmetric phase and stable fixed-point region with increasing the nonreciprocity for large nonlinear interactions. In addition, a linear self-trapping effect solely induced by the nonreciprocity have been revealed. We have shown the enhancement, collapse, and revival of self-trapping oscillations in the $\mathcal{PT}$-symmetric, $\mathcal{PT}$-broken, and reentrant $\mathcal{PT}$-symmetric phases, respectively. Considering that the tunable nonreciprocal hopping, double-well potential and interactions for atomic condensates have already been experimentally achieved, we expect that our predictions can be tested in future experiments with ultracold atoms. It would be interesting to study the full quantum dynamics in similar non-reciprocal systems in future work, as the classical correspondence of non-Hermitian quantum many-body dynamics is still far from being understood.

\acknowledgments{This work was supported by the National Natural Science
Foundation of China (Grants No. 12174126, No. 12104166, No. U1830111, No. 11947097, and No. 12047522), the Key-Area Research and Development Program of Guangdong Province (Grant No. 2019B030330001), the Science and Technology of Guangzhou (Grant No. 2019050001), and the Guangdong Basic and Applied Basic Research Foundation (Grants No. 2020A1515110290 and No. 2021A1515010315).}

\bibliography{reference}

\end{document}